\documentclass[a4paper, 12pt]{article}
\usepackage{amsmath}
\usepackage{amssymb}
\usepackage{epsfig}

\title{Central charges of wrapped M5-brane backgrounds}
\author{Jos\'e S\'anchez Loureda\footnote{email: j.m.sanchez-loureda@durham.ac.uk}
\quad and \quad
Douglas J. Smith\footnote{email: Douglas.Smith@durham.ac.uk} }

\begin{document}

\maketitle

\begin{center}

{\em Department of Mathematical Sciences,
University of Durham,
Science Laboratories,
South Rd,
Durham, DH1 3LE,
UK.}

\end{center}

\vspace{1.4cm}

\begin{abstract}

We study the central charges of the supersymmetry algebra of branes in backgrounds corresponding to wrapped M5-branes. In the case of M5-branes wrapping a holomorphic 2-cycle in $\mathbf{C}^{2}$, we find this allows for a supersymmetric M5-brane probe which is related to the M2-brane probe which describes the BPS spectra of the corresponding $\mathcal{N} = 2$ worldvolume gauge theory. For the case of M5-branes wrapping a holomorphic 2-cycle in $\mathbf{C}^{3}$, we find that the central charges allow for a supersymmetric M5-brane probe wrapping a Cayley calibrated 4-cycle, which has an intersecting BPS domain wall interpretation in the corresponding $\mathcal{N} = 1$ MQCD gauge theory. The domain wall is constructed explicitly as an M5-brane wrapping an associative 3-cycle. The tension is found to be the integral of a calibrating form. These wrapped M5-brane backgrounds provide a clear and interesting geometrical realisation of structure groups of M-theory vacua with fluxes.

\end{abstract}

\vspace{-20cm}
\begin{flushright}
DCPT-06/01 \\
\end{flushright}

\thispagestyle{empty}

\newpage

\setcounter{page}{1}

\section{Introduction}
\label{Introduction}

It was first established in~\cite{deAzcarraga:1989gm} that $p$-brane solitons in supersymmetric theories carry $p$-form charges that extend the spacetime supertranslation algebra. In particular these $p$-forms appear as central charges of the algebra, and since they are topological in nature, they depend only on the homology class of the configuration. This allows massive objects such as branes~\cite{Polchinski:1995df,Witten:1995ex,Hull:1994ys,Townsend:1995kk}, which carry the charge, to have supersymmetric ground states.

The analysis has been further generalised to include arbitrary supersymmetric backgrounds and worldvolume fields in eleven~\cite{Hackett-Jones:2003vz} and ten~\cite{Hackett-Jones:2004yi} dimensions. These topological charges of the algebra indicate what supersymmetric brane probes are possible in the chosen background, and also what possible relevance to the corresponding worldvolume gauge theory they might have.

Extending the analysis of~\cite{SanchezLoureda:2005ap}, we have chosen the supersymmetric backgrounds corresponding to M5-branes wrapped on holomorphic 2-cycles in $\mathbf{C}^{2}$~\cite{Fayyazuddin:1999zu,Fayyazuddin:2000em,Brinne:2000fh} and $\mathbf{C}^{3}$~\cite{Brinne:2000nf}. These correspond to completely localised intersecting brane configurations, describing a large class of Hanany-Witten type models when reduced dimensionally to Type IIA ten-dimensional string theory. The first case corresponds to an $\mathcal{N} = 2$ four-dimensional super Yang-Mills worldvolume theory~\cite{Witten:1997sc}, whilst the latter describes an $\mathcal{N} = 1$ MQCD gauge theory~\cite{Hori:1997ab,Witten:1997ep}.

An important element in obtaining the results presented here has been the G-structures inspired bilinear spinor formalism, which has already proved useful in understanding the general structure of supersymmetric solutions of supergravity theories~\cite{Gauntlett:2002fz,Gauntlett:2003wb}. In particular, certain $p$-forms built from the spinor fields which obey the Killing spinor equations for our chosen background are the basic building blocks of the central charges, along with the background flux and worldvolume fields. The classification of the structure groups of M-theory vacua with and without flux has benefited from this approach~\cite{Figueroa-O'Farrill:1999tx,Gillard:2004xq,Cariglia:2004ym}. We give a physical realisation of these structure groups from the allowed supersymmetric probes of the backgrounds under investigation.  

Geometrically, the M5-branes wrapping 2-cycles in $\mathbf{C}^{2}$ and $\mathbf{C}^{3}$ satisfy generalised calibrations. We find our results are consistent with the literature~\cite{SanchezLoureda:2005ap,Fayyazuddin:2005pm,Fayyazuddin:2005ds,Martelli:2003ki}. As we shall see, one of the results from the central charge calculation is that, for the case of an M5-brane wrapping a 2-cycle in $\mathbf{C}^{3}$, one can place a supersymmetric M5-brane probe wrapping a Cayley 4-cycle with respect to the null structure group $(SU(4) \ltimes {\mathbb{R}}^{8}) \times \mathbb{R}$, which can then be split to M5-branes wrapping an associative 3-cycle in relation to $(SU(3) \ltimes {\mathbb{R}}^{6}) \times {\mathbb{R}}^3$, and this is interpreted as intersecting BPS domain walls in the MQCD gauge theory (see~\cite{Fayyazuddin:1997ky,Gauntlett:2000ch,Gauntlett:2000yv} for instance). The case of an M5-brane wrapping a 2-cycle in $\mathbf{C}^{2}$ also admits M5-brane BPS-type probes, and a richer geometrical structure was uncovered.

The outline of this paper is as follows. In section~\ref{SolutionM5}, we quickly review the $\mathcal{N} = 2$ and $\mathcal{N} = 1$ supergravity solutions of localised brane intersections we shall be using for the analysis in the next sections. Section~\ref{SpinorBilinears} summarises the bilinear spinor formalism for our choice of background as well as discussing structure groups of M-theory vacua briefly. Section~\ref{CentralCharge} reviews the most general form of the M2 and M5-brane central charge. Section~\ref{N=2CentralCharge} contains the calculation of the $\mathcal{N} = 2$ M5-brane central charge, which contains the possibility of additional supersymmetric M5-brane probes by exhausting all the possible complex structures of the hyper-K\"{a}hler manifold in $\mathbf{C}^{2}$. These provides a physical illustration of the time-like M-theory structure groups, in particular the product group $SU(2) \times SU(3)$. Section~\ref{N=1CentralCharge} presents the results for the analogous $\mathcal{N} = 1$ M5-brane central charge, which contains the possibility of M5-branes wrapped on Cayley calibrated 4-cycles with respect to $(SU(4) \ltimes {\mathbb{R}}^{8}) \times \mathbb{R}$. These are interpreted as finite tension intersecting BPS domain walls in the MQCD gauge theory. These give a clear physical understanding of the null M-theory structure groups. Finally, the last section summarises the results obtained and discusses their implications.

\section{Supergravity solutions of M5-branes wrapped holomorphically in $\mathbf{C}^{2}$ and $\mathbf{C}^{3}$}
\label{SolutionM5}

\subsection{Review of the $\mathcal{N} = 2$ solution}

In this section we summarise briefly the eleven-dimensional $\mathcal{N} = 2$ supergravity solution of fully localised M5-brane intersections~\cite{Fayyazuddin:1999zu,Fayyazuddin:2000em,Brinne:2000fh}. Viewed from an M-theory perspective, this corresponds to an M5-brane with worldvolume $\mathbf{R^{(1,3)}} \times \Sigma$, where $\Sigma \subset \mathbf{C}^{2}$ is a Riemann surface. This is a holomorphic embedding which preserves $\mathcal{N} = 2$ (in $d=4$) supersymmetry. This brane configuration is related, in the appropriate near-horizon limit, to $\mathcal{N} = 2$ supersymmetric gauge theories by the $AdS/CFT$ correspondence~\cite{Maldacena:1997re}. 

This M-theory picture of an M5-brane wrapped on a holomorphic cycle of $\Sigma \subset \mathbf{C}^{2}$ has a ten-dimensional, Type IIA string theory interpretation. It describes a large class of Hanany-Witten~\cite{Hanany:1996ie} constructions. Generically, the Hanany-Witten setup involves D4-branes with worldvolume directions $01236$ ending on NS5-branes extended in the $012345$ directions. All the branes are located at $x^{8}=x^{9}=x^{10}=0$. We can define the complex coordinates $v=x^{4} + ix^{5}$ and $s=x^{6} + i x^{7}$, where $x^{7}$ is the eleventh dimension (a circle of radius $R$). This complex structure plays an important part in defining the complex manifold $\Sigma$ which the M5-brane wraps in the M-theory picture.

This Riemann surface $\Sigma$ is in fact the Seiberg-Witten curve for the gauge theory~\cite{Seiberg:1994rs}. The Seiberg-Witten differential also has an M-theory derivation~\cite{Fayyazuddin:1997by} (see~\cite{Elitzur:1997fh,Elitzur:1997hc} for a comprehensive review of these constructions). The BPS states correspond to minimal M2-branes whose boundary is on the M5-brane. The mass of the M2-brane gives the mass of the corresponding BPS-saturated state.

If we take the near-horizon limit of our brane setup, then the relevant scalings are defined such that $l_{P} \rightarrow 0$ with fixed $w = \frac{vR}{{l_{P}}^{3}}$ and $y = \frac{s}{R}$ (also, $t^{2} =  \frac{r}{{l_{P}}^{3}}$ with $r^2=\Sigma x_{i}^2$ $i=8,9,10$). 

Since the M5-brane is embedded holomorphically in the target space, we can define complex coordinates $F^2(w,y)$ and $G(w,y)$ that are holomorphic functions of $w,y$ and can be thought of as local co-ordinates transverse and parallel to the M5-brane. They also satisfy the constraint

\begin{equation}
\left({\partial}_{y}F^{2}\right) \left( {\partial}_{w} G \right) - \left({\partial}_{w}F^{2}\right) \left( {\partial}_{y} G \right) = 1.
\label{FGconstraint}
\end{equation}
The above equation is simply the statement that the Jacobian of the holomorphic co-ordinate transformation from $(w,y)$ to $(F^{2}, G)$ is equal to one. It is also the necessary condition for the metric

\begin{equation}
g_{M \bar{N}} \equiv 2 \left( {\partial}_{M} F^2 \right) \left( \overline{ {\partial}_{N} F^2 } \right) g + 1/2 \left( {\partial}_{M} G \right) \left( \overline{ {\partial}_{N} G } \right)
\end{equation}
to have determinant $g$. Here $g$ is a harmonic function in the five-dimensional transverse space with radial co-ordinate $\tilde{r} \equiv \sqrt{t^4 + \left\vert F \right\vert ^4}$ so that $g= \frac {\pi N}{8 {\tilde{r}}^3 }$.

The spacetime metric for an M5-brane that wraps a holomorphic 2-cycle in $\mathbf{C}^{2}$ is naively of the form $\mathbf{R^{(1,3)}} \times M_{4} \times \mathbf{R^{(3)}}$ and given by:

\begin{eqnarray}
ds^2 &=& H^{-1/3} {dx^2}_{(1,3)} + 2 H^{-1/3} g_{M \bar{N}} dz^{M}dz^{\bar{N}} + H^{2/3} {dx^2}_{(3)} \\
\label{FSspacetimeC2}
H &=& 4g ,
\end{eqnarray}
where $M_{4}$ is a 2 complex-dimensional hyper-K\"{a}hler manifold. We denote with Greek letters $\alpha,\beta,\gamma$ the totally transverse directions $8,9,10$, and capital letters $M,N$ for the complex co-ordinates $F^2,G$. 

Defining the Hermitian two-form $\omega_{G}= i G_{M\bar{N}} dz^{M} \wedge dz^{\bar{N}}$ (where we have rescaled the metric $G_{M\bar{N}}=H^{-1/3}g_{M\bar{N}}$), the spacetime metric is found to satisfy a (warped) K\"{a}hler calibration constraint

\begin{equation}
d_{\mathbf{C}^{2}} \left[ H^{1/3} \omega_{G} \right] = 0,
\label{C2constraint}
\end{equation}
with the derivative understood to be acting on the complex submanifold. This constraint is something which can be seen by generalised calibration arguments or otherwise~\cite{Fayyazuddin:2000em,Fayyazuddin:2005pm}. 

The non-vanishing components of the four-form field strength are:

\begin{eqnarray}
F_{M\bar{N}\alpha\beta} & = & 2i \epsilon_{\alpha\beta\gamma}\partial_{\gamma}g_{M\bar{N}} \nonumber \\
F_{M89(10)} & = & - i \partial_{M} H \nonumber \\
F_{\bar{N}89(10)} & = & i \partial_{\bar{N}} H 
\label{C2FieldStrength}
\end{eqnarray}
These can be calculated simply once we note that the calibrating form $\Phi$ of the M5-brane is equal to its volume form, since it is a supersymmetric object. From the metric, this is can be seen to be 

\begin{eqnarray}
\Phi & = & - i H^{-2/3} G_{M\bar{N}} dt \wedge dX^1 \wedge dX^2 \wedge dX^3 \wedge dz^{M} \wedge dz^{\bar{N}} \nonumber \\
& = & dV_{0123} \wedge \omega_{G}.
\label{BM5Calibration}
\end{eqnarray}
Taking the Hodge dual of the exterior derivative and then using the constraint~(\ref{C2constraint}) gives the result~(\ref{C2FieldStrength}), since $F_{4} = \ast F_{7} = \ast d \Phi$. 

\subsection{Review of the $\mathcal{N} = 1$ solution}
\label{N1SYM}

The previous construction, which describes the eleven-dimensional supergravity dual of $\mathcal{N}=2$ field theories as the near-horizon limit of an M5-brane wrapped on a Riemann surface $\Sigma \subset \mathbf{C}^2$, has been generalised to $\mathcal{N}=1$~\cite{Brinne:2000nf}. In particular, the eleven dimensional supergravity dual of certain $\mathcal{N}=1$ field theories, (so-called MQCD theories~\cite{Hori:1997ab,Witten:1997ep}), is given by the near-horizon limit of an M5-brane wrapped on a Riemann surface $\Lambda \subset \mathbf{C}^3$. MQCD is then the quantum field theory living on the $\mathbf{R^{(1,3)}}$ part of a five-brane with world-volume $\mathbf{R^{(1,3)}} \times \Lambda$. 

An M5-brane wrapped on a holomorphic 2-cycle in $\mathbf{C}^3$ also has a ten-dimensional Type IIA string theory interpretation. It is similar to the Hanany-Witten construction for the $\mathcal{N}=2$ solution, except now one of the NS5-branes has been rotated from the $45$ plane to the $89$ plane. This rotation corresponds to turning on a mass for the adjoint scalar in the $\mathcal{N}=2$ vector multiplet, breaking the supersymmetry down to $\mathcal{N}=1$.

We can define the complex coordinates $z^1 = x^4 + ix^5$, $z^2 = x^6 + ix^7$ and $z^3 = x^8 + ix^9$, with $y=x^{(10)}$ as the only totally transverse direction. 

The spacetime metric for an M5-brane that wraps a holomorphic 2-cycle in $\mathbf{C}^{3}$ is of the form $\mathbf{R^{(1,3)}} \times M_{6} \times \mathbf{R}$ and given by:

\begin{eqnarray}
ds^2 &=& H^{-1/3} {dx^2}_{(1,3)} + 2 H^{1/6} g_{M \bar{N}} dz^{M}dz^{\bar{N}} + H^{2/3} dy^2  
\label{FSspacetimeC3} \\
\det g &=& H , \nonumber 
\end{eqnarray}
with $M_{6}$ being a 3 complex-dimensional Hermitian manifold. The metric tensor $g_{M\bar{N}}$ is Hermitian, and has an associated Hermitian 2-form

\begin{equation}
\omega = i g_{M\bar{N}} dz^M\wedge dz^{\bar{N}},
\end{equation}
which is useful in expressing the field strength $F$ in a more elegant form

\begin{equation}
F = \partial_{y}(\omega\wedge\omega)-i\partial(H^{1/2}\omega)\wedge dy + i\bar{\partial}(H^{1/2}\omega)\wedge dy. 
\label{C3FieldStrength}
\end{equation}
These non-vanishing components can again be worked out easily from noticing that the calibrating form $\Phi$ of the M5-brane is equal to its volume form, since it is a supersymmetric object. From the metric, this is can be seen to be 

\begin{eqnarray}
\Phi & = & i H^{-1/2} g_{M\bar{N}} dt \wedge dX^1 \wedge dX^2 \wedge dX^3 \wedge dz^{M} \wedge dz^{\bar{N}} \nonumber \\
& = & dV_{0123} \wedge \omega .
\label{CalboundC3}
\end{eqnarray}
Taking the Hodge dual of the exterior derivative of this form gives the result~(\ref{C3FieldStrength}), since $F_{4} = \ast F_{7} = \ast d \Phi$.

The spacetime metric is found to satisfy a co-K\"{a}hler calibration constraint

\begin{equation}
d_{\mathbf{C}^3}(\omega\wedge\omega) = 0 \; = \; d_{\mathbf{C}^3}\ast\omega,
\label{C3constraint}
\end{equation}
where the exterior derivative and Hodge duality operation naturally take place in the $\mathbf{C}^3$ submanifold. This can be seen from generalised calibration arguments~\cite{Brinne:2000nf,Fayyazuddin:2005ds}, for example.

\section{Bilinear spinor formalism}
\label{SpinorBilinears}

A brief overview of the spinor formalism we shall be using throughout is useful to set our definitions and conventions. We use the notation in~\cite{Smith:2002wn}, with $\Gamma_{M}$ for the spacetime Dirac gamma-matrices and $\hat{\Gamma}_{m}$ for the tangent-space gamma-matrices. These are related by the vielbein $e^{m}_{M}$ such that

\begin{equation}
g_{MN}=e^{m}_{M}e^{n}_{N} \eta_{mn} \; , \; \Gamma_{M}=e^{n}_{M}\hat{\Gamma}_{m} \; , \; \left\{ \Gamma_{M} , \Gamma_{N} \right\} =2g_{MN} \; , \; \left\{ \hat{\Gamma}_{m} , \hat{\Gamma}_{n} \right\} =2\eta_{mn} .
\end{equation}
The number of supersymmetries preserved by a p-brane configuration is given by the number of spinors $\epsilon$ which satisfy the equation  

\begin{equation}
\hat{\Gamma}\epsilon = \epsilon
\end{equation}
where we have the definitions $\hat{\Gamma} = \frac{1}{p!} \epsilon^{\alpha_{1}\ldots\alpha_{p}} \Gamma_{M_{1}\ldots M_{p}} \partial_{\alpha_{1}}X^{M_{1}}\ldots \partial_{\alpha_{p}}X^{M_{p}}$ and \\
$\Gamma_{M_{1}\ldots M_{p}} = \frac{1}{p!} \Gamma_{\left[ M_{1}\ldots \right.} \Gamma_{\left. M_{p}\right]}$. The $X^M$ is the embedding of the p-brane in the background geometry and the $\alpha_{i}$ denote the worldvolume coordinates. 

If we consider for definiteness our $\mathcal{N} = 1$ example, we know that the asymptotic form of the embedding (i.e.\ the Type IIA Hanany-Witten type model) should be three sets of orthogonally intersecting M5-branes with worldvolume directions $012345$, $012367$ and $012389$. This corresponds to the projection conditions

\begin{eqnarray}
\hat{\Gamma}_{012345} \epsilon &=& \epsilon \nonumber \\
\hat{\Gamma}_{012367} \epsilon &=& \epsilon \nonumber \\
\hat{\Gamma}_{012389} \epsilon &=& \epsilon .
\end{eqnarray}

There are, however, alternative ways to write these conditions. For example, the above projection conditions imply that

\begin{equation}
\hat{\Gamma}_{4567}\epsilon =  \hat{\Gamma}_{6789}\epsilon = - \epsilon .
\end{equation}
Since we are going to be working with holomorphic cycles in a target space with a complex structure, it makes sense to re-write the Clifford algebra in terms of complex coordinates. If we now define complex co-ordinates 
\begin{eqnarray}
z^{1} &=& x^{4}+ix^{5} \nonumber \\
z^{2} &=& x^6 +ix^7 \nonumber \\
z^{3} &=& x^8 + ix^9 
\end{eqnarray}
and complex gamma matrices 

\begin{equation}
\hat{\Gamma}_{z^{j}} = \frac{1}{2} \left( \hat{\Gamma}_{x^{2j+2}} - i \hat{\Gamma}_{x^{2j+3}} \right),
\end{equation}
then we can concisely express the above relations as

\begin{equation}
\hat{\Gamma}_{0123a\bar{b}}\epsilon = i \delta_{a\bar{b}}\epsilon,
\label{ProjM5}
\end{equation}
with $\delta_{a\bar{b}}$ as the tangent-space metric with $\delta_{1\bar{1}}=\frac{1}{2}$ in our conventions. This restriction on $\epsilon$ means that the solution will preserve $\frac{1}{8}$ of the supersymmetry (or equivalently, four supercharges), which corresponds to $\mathcal{N}=1$ in four dimensions. We use conventions where $ds^2 = 2 g_{M\bar{N}} dz^M dz^{\bar{N}} = 2 \delta_{a\bar{b}} e^{a}_{M} \left( \overline{e^{b}_{N}} \right) dz^M dz^{\bar{N}}$ for complex Hermitian metrics.

It is useful to realise that the spinor $\epsilon(x)$ which satisfies the Killing spinor equation for preservation of supersymmetry can be reconstructed (up to a sign) from the following one, two and five-forms built from spinor bilinears:

\begin{eqnarray}
K_{M} & = &   \overline{\epsilon}\Gamma_{M}\epsilon \nonumber \\
\Omega_{MN} & = & \overline{\epsilon}\Gamma_{MN}\epsilon   \nonumber \\
\Sigma_{MNPQR} & = & \overline{\epsilon}\Gamma_{MNPQR}\epsilon .
\label{Onetwofiveforms}
\end{eqnarray}
One can check that the zero, three and four-forms built in a similar way vanish identically. Furthermore, as recent work on G-structures~\cite{Gauntlett:2002fz,Gauntlett:2003wb} and related ideas has emphasised, if we start with an eleven-dimensional geometry with a $Spin(10,1)$ structure and assume that we have a globally defined spinor, then, at a point, the isotropy group of the spinor is known to be either $SU(5)$ or $(Spin(7) \ltimes {\mathbb{R}}^{8}) \times \mathbb{R}$ depending on whether $K$ is time-like or null, respectively. There are also differential and algebraic relations that relate these forms to themselves and the background flux which have proved to be very useful in constructing supergravity solutions, but we shall not need them here.

Specialising to our backgrounds in question, we find that the case of the M5-brane wrapping a holomorphic cycle in $\mathbf{C}^{2}$ corresponds to having a time-like $K$, whereas the case of the M5-brane wrapping a holomorphic cycle in $\mathbf{C}^{3}$ will contain a null vector $K$. We shall construct these forms explicitly in the following sections. Once we have constructed the central charges, they shall provide a clear geometrical understanding of the supersymmetric brane probe moduli space.

\subsection{Structure groups of M-theory vacua}

The isometries of M-theory vacua with no flux have been known for some time now, for the cases where there exists either a time-like or a null globally defined Killing spinor. A classification of the holonomy groups of M-theory vacua preserving various fractions of supersymmetry also exists~\cite{Figueroa-O'Farrill:1999tx}. It is well known that, in general, the holonomy of the background geometry is reduced once fluxes (or equivalently a non-trivial four-form field strength) are turned on. There is in general a torsion which modifies the usual connection on the manifold, thus giving it a reduced group structure. In general, a spacetime has holonomy G if it admits a torsion-free G-structure. A G-structure on a spacetime $M$ can be defined as a principal sub-bundle of the frame bundle of $M$. So the deviation from holonomy of a supersymmetric spacetime with fluxes is encoded in the intrinsic torsion of the G-structure. Classifications of structure groups preserving different fractions of supersymmetry in eleven dimensions with flux have shown that more isometry subgroups are possible than in the no flux case. Physically, the reduced group structure arises due to the backreaction of the branes on the geometry. 

As we shall see in the following sections, the M5-brane probes allowed by the central charges have a worldvolume geometry consistent with the time-like and null M-theory structure groups. For the case of time-like structure groups, there is to our knowledge only a partial classification of structure groups arising as subgroups of $SU(5)$ in eleven dimensions~\cite{Gillard:2004xq}. However, the holonomy groups have been classified~\cite{Figueroa-O'Farrill:1999tx}. Since these will still be present, as reduced structure groups, once fluxes are included, they shall be enough for our purposes. These will correspond to the case of an M5-brane wrapping a 2-cycle in $\mathbf{C}^{2}$, and the various probes the central charges allow. For the case of null structure groups arising as subgroups of $(Spin(7) \ltimes {\mathbb{R}}^{8}) \times \mathbb{R}$, the complete classification has been done~\cite{Cariglia:2004ym}. This will fit in nicely with the physical picture revealed by the central charge calculation of M5-brane probes of the background of M5-branes wrapping a 2-cycle in $\mathbf{C}^{3}$. 

\section{Central charges of eleven-dimensional supergravity backgrounds}
\label{CentralCharge}

As we shall see, possible supersymmetric brane probes allowed by the background geometry are determined by the central charges of the brane configuration in question. In general these charges, being topological in nature, provide a clear picture of allowed supersymmetric objects in that particular background, and possible field theory interpretations.

The spacetime superalgebra we are considering here is the general eleven-dimensional super-Poincar\'e algebra coupled to a five-brane and non-zero background flux and worldvolume fields~\cite{Hackett-Jones:2003vz}. In its most general form, valid for either time-like or null $K$, the supersymmetry algebra on the worldvolume of a probe M5-brane becomes:

\begin{equation}
2\left(\epsilon Q\right)^2 = \int K^{M}P_{M} \pm \int \left(  \imath_{K}C + \Sigma + (A + dB) \wedge (\Omega + \imath_{K}A) - \frac{1}{2}A\wedge \imath_{K}A \right).
\end{equation} 
We have denoted by $C$ the background six-form potential and $A$ denotes the electric three-form potential, related by $dC = \ast dA + \frac{1}{2} A \wedge F$. We also include the non-zero worldvolume two-form gauge field $B$. Since $\left(\epsilon Q\right)^2 \geq 0$, this leads to a BPS type bound on the energy/momentum of the M5-brane,

\begin{equation}
\int K^{M}P_{M} \geq \mp \int \left(  \imath_{K}C + \Sigma + (A + dB) \wedge (\Omega + \imath_{K}A) - \frac{1}{2}A\wedge \imath_{K}A \right).
\end{equation} 

From properties of our construction, such as the fact that $K$ is Killing and $\mathcal{L}_{K} F = 0$, one can check that the right hand side of the inequality is the integral of a closed form and thus represents a topological charge. The existence of such a closed form also provides examples of generalised calibration forms for arbitrary supersymmetric backgrounds.

The analogous central charges for the supersymmetry algebra on the worldvolume of a probe M2-brane are given by

\begin{equation}
\int K^{M}P_{M} \geq \mp \int \left(  \Omega + \imath_{K} A  \right).
\end{equation}

\section{Central charges of an M5-brane wrapping a holomorphic 2-cycle in $\mathbf{C}^2$}
\label{N=2CentralCharge}

In this section we proceed to construct the central charges of probe M5-branes for the first background we are examining, M5-branes wrapped on a holomorphic 2-cycle in $\mathbf{C}^2$, which is the supergravity dual of $\mathcal{N}=2$ supersymmetric gauge theories in four dimensions.

As previously mentioned, this case corresponds to a background geometry with a time-like vector $K$. Since our background preserves eight supercharges, the original $SU(5)$ isotropy group will be broken down, reflecting the reduced isometries of our background. These can be described by the projection conditions that our configuration satisfies. We shall discuss the structure groups that arise as a consequence of this and its implications at the end of the section.

If we decompose the Hermitian metric $g_{M\bar{N}}$ into tangent space zweibeins such that 

\begin{eqnarray}
e^{1}_{M} &=& H^{1/2} \left(\partial_{M}F^2\right) \nonumber \\
e^{2}_{N} &=& \left(\partial_{N}G\right) ,
\end{eqnarray}
then the complex structure $J$ in which the M5-brane is embedded holomorphically is given by 

\begin{eqnarray}
dZ^1 & = & Re \left( e^{1}_{M} dz^{M} \right) + i Im \left(  e^{1}_{M} dz^{M} \right) \nonumber \\
dZ^2 & = & Re \left( e^{2}_{M} dz^{M} \right) + i Im \left(  e^{2}_{M} dz^{M} \right) .
\end{eqnarray}
The projection condition this M5-brane satisfies is given by 

\begin{equation}
\hat{\Gamma}_{0123a\bar{b}}\epsilon = i \delta_{a\bar{b}}\epsilon,
\label{ProjBM51}
\end{equation}
with $a,b=z^1,z^2$. It is also well known that this background admits a supersymmetric M2-brane probe which is a BPS state of the worldvolume theory of the M5-brane. Since the complex submanifold in which we are embedding this brane is actually hyper-K\"{a}hler (as are all two complex-dimensional Ricci-flat K\"{a}hler manifolds), this geometry admits a family of inequivalent complex structures parametrised by a two-sphere $S^2$, with $SU(2)$ commutation relations between them. Also, in four dimensions, the hyper-K\"{a}hler manifold should therefore admit a covariantly constant holomorphic two-form.

In order to ensure that the two-brane ends on the five-brane, we shall need to wrap the M2-brane on a holomorphic cycle with respect to a complex structure $J^\prime$ which is orthogonal to the complex structure $J$ in which the background M5-brane was embedded holomorphically. Given a complex structure $J$, the set of such $J^\prime$ for a hyper-K\"{a}hler manifold is parametrised by an $S^1$ that actually corresponds to the phase of the central charge of the BPS saturated state~\cite{Henningson:1997hy}.

In terms of the M5-brane holomorphic coordinates, the projection condition for the M2-brane can be written

\begin{equation}
\mathcal{P}\epsilon = \left(e^{i\phi}\hat{\Gamma}_{0ab} + e^{-i\phi}\hat{\Gamma}_{0\bar{a}\bar{b}}\right)\epsilon = \epsilon .
\label{ProjpM2}
\end{equation}
We have included the arbitrary phase $\phi$ for generality, and $a=z^1 , b=z^2$. We note that the linear combination of holomorphic and anti-holomorphic projection conditions do indeed insure it is an Hermitian projector with $\mathcal{P}^2 = 1$. This additional constraint cuts the number of supersymmetries by half (leaving four real supersymmetries), confirming that the M2-brane is a BPS object of the M5-brane worldvolume gauge theory. For more details see~\cite{SanchezLoureda:2005ap}.

The next thing to notice is that using the identity $\hat{\Gamma}_{0123456789(10)}\equiv 1$ we can show that our projections actually allow for another M5-brane that does not break any further supersymmetries. We find it wraps a holomorphic cycle with respect to a complex structure $J^{\prime\prime}$ which is orthogonal to both the previous cases and exhausts the three independent complex structures of the hyper-K\"{a}hler manifold they are embedded in. In addition, this M5-brane has a worldvolume extension along the $89(10)$ space. Explicitly, the projection condition for this ``hidden" M5-brane is 

\begin{equation}
\left(e^{i(\phi+\pi /2)}\hat{\Gamma}_{0ab} + e^{-i(\phi + \pi /2)}\hat{\Gamma}_{0\bar{a}\bar{b}}\right)\hat{\Gamma}_{89(10)}\epsilon = \epsilon
\label{ProjhM5}
\end{equation}
Finally, there is another projection condition which is compatible and commutes with these three (and therefore does not break supersymmetry any further). This can best be expressed by defining the complex coordinates

\begin{eqnarray}
\xi^{1} &=& e^1 + i e^{(10)} \nonumber \\
\xi^{2} &=& e^2 + i e^9 \nonumber \\
\xi^{3} &=& e^3 + i e^8
\end{eqnarray}
which allows us to write the projection condition in the following way:

\begin{equation}
\hat{\Gamma}_{0\xi^i \xi^{\bar{j}}}\epsilon = i \delta_{i\bar{j}} \epsilon ,
\end{equation}
with $i,j=\xi^1 ,\xi^2 ,\xi^3$. As before we have $\delta_{1\bar{1}} = 1/2$. That this additional complex submanifold is compatible with all our earlier projections somehow reflects that the original, larger $SU(5)$ isometry group has been broken down by the appearance of branes in the geometry. This is consistent with the known structure groups of our background, as we shall discuss. We can show this more clearly by combining the preceding constraints to build the following projection, which shows some of that residual structure:

\begin{equation}
\left(\hat{\Gamma}_{0\xi^1 \xi^2 \xi^3 a\bar{b}} + \hat{\Gamma}_{0\bar{\xi}^1 \bar{\xi}^2 \bar{\xi}^3 a\bar{b}}\right)\epsilon = \frac{i}{4} \delta_{a\bar{b}} \epsilon .
\label{ProjBGenM5}
\end{equation}
One can check that this projection condition includes the original background M5-brane projection~(\ref{ProjBM51}). It is also reminiscent of an M5-brane wrapping a special Lagrangian 3-cycle in one manifold $\tilde{M}_{6}$, and a holomorphic 2-cycle in the hyper-K\"{a}hler manifold $M_{4}$. We shall expand on this geometrical statement a little later, where we shall see that $\tilde{M}_{6}$ will turn out to be a complex manifold. We discuss possible field theory interpretations at the end of the section.

These conditions then complete the set of independent, commuting projections and thus determine a unique spinor up to scale. The scale of the spinor, which we will use shortly to calculate the forms $K,\Omega$ and $\Sigma$, is given by fixing $\epsilon^{\dagger}\epsilon = \Delta$. Using the fact that $K$ is a Killing vector of our background, we found that $\Delta = \sqrt{-g_{00}}$. We can now proceed to calculate the non-trivial components of each form. A quick calculation reveals that $K_{i}=0$ for $\{i=1,2,3,a,\bar{b},8,9,(10)\}$, since, for example,

$$
\begin{array}{ccccccc}
\hat{K}_{1} & = & \overline{\epsilon} \hat{\Gamma}_{1} \epsilon & = &  \overline{\epsilon} \hat{\Gamma}_{1} \left( e^{i\phi}\hat{\Gamma}_{0ab} + e^{-i\phi}\hat{\Gamma}_{0\bar{a}\bar{b}} \right) \epsilon & = & - \overline{\epsilon} \left( e^{i\phi}\hat{\Gamma}_{01ab} + e^{-i\phi}\hat{\Gamma}_{01\bar{a}\bar{b}} \right) \epsilon \\
& = & 0 
\end{array}
$$
where in the second step we have used the M2-brane projection condition and in the last step the fact that four-forms constructed in this way vanish identically.

After some work, the resulting forms turn out to be:

\begin{eqnarray}
K & = & -H^{-1/3} dt \\
\Omega &=& 1/2 \left( e^{-i\phi} dz^1 \wedge dz^2 + e^{i\phi} d\bar{z}^1 \wedge d\bar{z}^2 \right) + i H^{-1/6} \delta_{i\bar{j}} \; \xi^i \wedge \xi^{\bar{j}} \\
\Sigma &=&  H/2 \left( e^{-i(\phi + \pi /2)} dz^1 \wedge dz^2 + e^{i(\phi + \pi /2)} d\bar{z}^1 \wedge d\bar{z}^2 \right) \wedge dx^8 \wedge dx^9 \wedge dx^{(10)} \nonumber \\
& & + i H^{-1/2} g_{M\bar{N}} \; Re \left( \xi^1 \wedge \xi^2 \wedge \xi^3 \right) \wedge dz^M \wedge dz^{\bar{N}} \nonumber \\ 
& & - 1/4 \; dt \wedge dz^1 \wedge d\bar{z}^1 \wedge dz^2 \wedge d\bar{z}^2 \nonumber \\
& & - H^{-1/3} \delta_{i\bar{j}} \delta_{k\bar{l}} \; dt \wedge \xi^i \wedge \xi^{\bar{j}} \wedge \xi^k \wedge \xi^{\bar{l}} . 
\end{eqnarray}
%
%
%
%

%
The last two terms above do not play a role for possible static probe branes but we can consider taking the Hodge dual. These could be part of the D6-brane central charge in the Type IIA ten-dimensional picture, which is only geometry in eleven dimensions. 

Now we recall that in order to calculate the central charge we also need the six and three-form potentials for our particular background. The six-form potential can be ascertained easily if we remember that the background M5-brane satisfies a generalised calibration~(\ref{BM5Calibration}) which, by virtue of the BPS supersymmetry condition, is gauge equivalent to the spacetime gauge potential under which it is charged. We proceed by working with the asymptotically flat background co-ordinates $z^i$ before taking the near-horizon limit. Taking into account that the potential vanishes at spatial infinity, and taking the contraction with respect to $K$, we can conclude that

\begin{equation}
\imath_{K}C = - i (H^{-1}g_{M\bar{N}} - \delta_{M\bar{N}}) \; dx^1 \wedge dx^2 \wedge dx^3 \wedge dz^M \wedge dz^{\bar{N}}  
\label{C2Potential}
\end{equation} 
The only thing that remains is to define the three-form $A$ since, as in this case it is a magnetic potential, it is not globally well defined. The natural solution is to define the integral of $A\wedge\Omega$ over the spatial worldvolume of the brane such that 

\begin{equation} 
\int_{M5} F\wedge\Omega = \int_{\partial M5} A\wedge\Omega .
\end{equation}
Direct calculation reveals that the product is given by

\begin{eqnarray}
\Omega\wedge F & = & -1/2 \; dH \wedge \left(e^{-i(\phi + \pi /2)} dz^1 \wedge dz^2 + e^{i(\phi + \pi /2)} d\bar{z}^1 \wedge d\bar{z}^2 \right) \wedge \nonumber \\
& & dx^8 \wedge dx^9 \wedge dx^{(10)} \nonumber \\
& & + (dx^1 \wedge dx^{(10)} + dx^2 \wedge dx^9 + dx^3 \wedge dx^8 ) \wedge \nonumber \\
& & i \epsilon_{\alpha\beta\gamma}\partial_{\gamma}g_{M\bar{N}} dz^{M} \wedge dz^{\bar{N}} \wedge dx^{\alpha} \wedge dx^{\beta} 
\end{eqnarray}
so we can deduce that the magnetic potential $A$ can be defined,

\begin{eqnarray}
\Omega\wedge A &=&  -1/2 \; (H-1) \left(e^{-i(\phi + \pi /2)} dz^1 \wedge dz^2 + e^{i(\phi + \pi /2)} d\bar{z}^1 \wedge d\bar{z}^2 \right) \wedge \nonumber \\
& & dx^8 \wedge dx^9 \wedge dx^{(10)} \nonumber \\
& & -i \left( dx^1 \wedge dx^8 \wedge dx^9 + dx^3 \wedge dx^9 \wedge dx^{(10)} \right. \nonumber \\
& & \left. - dx^2 \wedge dx^8 \wedge dx^{(10)} \right) \wedge (g_{M\bar{N}} - \delta_{M\bar{N}}) dz^M \wedge dz^{\bar{N}}.
\end{eqnarray}
%
%
We can see straight away from this expression that the contraction $\imath_{K}A$ vanishes. 
We note that we can define the (1,1)-form $J= \frac{i}{2} ( e^1 \wedge e^{\bar{1}} + e^2 \wedge e^{\bar{2}})$ on the $M_{4}$ manifold and also define a (1,1)-form on $\tilde{M}_{6}$ by $\tilde{J} =\frac{i}{2} (\xi^1 \wedge \xi^{\bar{1}} + \xi^2 \wedge \xi^{\bar{2}} + \xi^3 \wedge \xi^{\bar{3}})$. The justification for this will be discussed as the end of the section in terms of structure groups. We may also define the flat space K\"{a}hler form $J_{f}=i \delta_{a\bar{b}}dz^a \wedge dz^{\bar{b}}$ on $M_{4}$ and also the flat holomorphic three-form $\tilde{\Psi}_{3f}= (dx^1 +idx^{(10)} ) \wedge (dx^2 +idx^9 ) \wedge (dx^3 +idx^8 )$ and the flat K\"{a}hler form $\tilde{J}_{f}=iH^{-1/6}\delta_{i\bar{j}} \xi^i \wedge \xi^{\bar{j}} = H^{-1/6}\tilde{J}$ on $\tilde{M}_{6}$.

Taking this into account and assembling all the terms, changing $w,y$ to $F^2 ,G$ where appropriate, we get that the central charges on an M5-brane probe of an M5-brane wrapping a holomorphic 2-cycle in $\mathbf{C}^2$ is given by

\begin{eqnarray}
\int K^{M}P_{M}  \; \geq  &\mp & \int \left(  \imath_{K}C + \Sigma + (A + dB) \wedge \Omega \right) \nonumber \\
 = & \mp & \int \Bigl( Re (\tilde{\Psi}_{3f}) \wedge J_{f} \Bigr. \nonumber \\
& & \left. + \frac{1}{2} \left( e^{-i(\phi + \pi /2)} dF^2 \wedge dG + e^{i(\phi + \pi /2)} d\bar{F}^2 \wedge d\bar{G} \right) \wedge dx^8 \wedge dx^9 \wedge dx^{(10)} \right. \nonumber \\
& & +  \; dt \wedge  \; J_{f} \wedge J_{f}  \nonumber \\
& &  +  \; dt \wedge   \; \tilde{J}_{f} \wedge \tilde{J}_{f} \nonumber \\
& & \Bigl. + dB \wedge \Omega \Bigr) .  
\end{eqnarray}
We note that the supersymmetry algebra is unaltered from flat space for a suitable choice of co-ordinates, of the same form as Eq. $(8)$ in~\cite{Hackett-Jones:2003vz}. The first term indicates the obvious possibility that an M5-brane probe which is parallel to the background M5-brane is an allowed supersymmetric probe. There are also other possibilities such as an M5-brane with spatial embedding $2458(10)$ for example. Depending on the boundary conditions, these may have field theory interpretations. 

In addition, the second term allows for an M5-brane which is embedded holomorphically in the hyper-K\"{a}hler manifold with respect to $J^{\prime\prime}$ and extended along $89(10)$. 

The possible holomorphy conditions on the pullback onto the probe branes are given by

\begin{eqnarray}
H^{1/2} \frac{\partial F^2}{\partial \sigma^1} &=& e^{i\phi} \frac{\partial \bar{G}}{\partial \sigma^2} \nonumber \\
H^{1/2} \frac{\partial F^2}{\partial \sigma^2} &=& - e^{i\phi} \frac{\partial \bar{G}}{\partial \sigma^1} ,
\label{HolRestraintsGenFG}
\end{eqnarray}
where we have defined $\sigma = \sigma^1 + i \sigma^2$ to be the complex coordinate on the probe worldvolume . We have chosen the different complex structures $J, J^{\prime}$ and $J^{\prime\prime}$ so they are specified by $\phi = 0, \pi /2 , -\pi / 2$, respectively. 

This probe would be related to the M2-brane probe which gives the masses of BPS states in the worldvolume gauge theory of the background M5-brane under appropriate boundary conditions. In our present notation, we recall that the central charge of the probe M2-brane would be given by

\begin{equation}
\mp \int \left(   1/2 \left( e^{-i\phi} dF^2 \wedge dG + e^{i\phi} d\bar{F}^2 \wedge d\bar{G} \right) + i H^{-1/6} \tilde{J}  \right) .
\end{equation}
If we look at the first term above, the difference between the two would be a volume modulus of the $89(10)$ space and also a rotation of the complex structure. Depending on the boundary conditions, this extra volume modulus could well be finite, like in the vortex case.

The second term above $i H^{-1/6} \tilde{J}$ is a calibration form of the $\tilde{M}_{6}$ manifold and denotes the possibility of co-dimension two objects on the worldvolume theory. These turn out to correspond to BPS vortices, something that can be pictured in terms of Hanany-Witten models~\cite{Tong:2005un}. Imposing a suitable boundary condition along the totally transverse directions $89(10)$ could allow these vortices to have finite tension.

The last couple of terms of the M5-brane central charge cannot be pulled back consistently to a static probe brane so we may consider taking the Hodge dual of them. We note, however, that the quantities in the brackets $(H^{-1/3} \; J \wedge J)$ and $(H^{-1/3}  \; \tilde{J} \wedge \tilde{J})$ are calibrating forms. Taking the dual then gives terms that would contribute to the Type IIA ten-dimensional central charges for the D6-brane, which in M-theory is given by pure geometry. These terms can be re-written

\begin{eqnarray}
\Sigma_{\ast(0JJ)} &=& - \ast \left( H^{-1/6} \; e^0 \wedge J \wedge J \right) \nonumber \\
& & - \ast \left( H^{-1/6}  \; e^0 \wedge \tilde{J} \wedge \tilde{J} \right) \nonumber \\
&=& - H^{-1/6} \; \tilde{J} \wedge \tilde{J} \wedge \tilde{J} \nonumber \\
& & - H^{-1/6} \; \tilde{J} \wedge J \wedge J .
\end{eqnarray}
We see that there are only terms which give the volumes of $\tilde{M}_{6}$ and $M_{4}$ respectively. Since we are compactifying along the M-theory circle (contained in $M_{4}$), only the first term would contribute to the D6-brane central charge.

To make sense of the various forms we have defined and arguments about the manifold $\tilde{M}_{6}$ we have to talk about structure groups. To our knowledge, there is no complete classification for eleven dimensional backgrounds with a time-like Killing spinor and flux, of which the brane configuration studied in this section is an example. However, from what is known from holonomy groups of M-theory vacua with no flux~\cite{Figueroa-O'Farrill:1999tx}, we can deduce the structure group for our configuration. It is easy to see that, for the case of our M5-brane wrapped on a 2-cycle in $\mathbf{C}^{2}$ with the BPS M2-brane probe ending on it and wrapped on a 2-cycle in a different complex structure, we preserve $\frac{1}{8}$ of supersymmetry. Backgrounds preserving this fraction of supersymmetry allow for three possible structure groups, but we can deduce that the appropriate one for our case is that we should have an overall $SU(2) \times SU(3)$ structure. This fits in with the known $M_{4}$ manifold typical of these spacetimes, whilst uncovering the $\tilde{M}_{6}$ manifold which was hinted at by the allowed projection conditions. The fact that $\tilde{M}_{6}$ has an $SU(3)$ structure then means that, although it is no longer Calabi-Yau (since this requires it to have $SU(3)$ holonomy), it is still a complex manifold. We can therefore use a similar technique to~\cite{Fayyazuddin:2005ds}, to recover calibration forms for this manifold and for the product manifold $M_{4} \times \tilde{M}_{6}$.

As we shall see in the next section, where the classification of eleven dimensional backgrounds with flux and a null Killing spinor has been done, these structure groups provide an elegant illustration of the transitions of the wrapped M5-brane's worldvolume, which give rise to intersecting BPS domain wall configurations on its worldvolume theory.

Another way to look at our background is to say that it is globally of the form $\mathbf{R} \times \mathbf{M}_{4} \times \mathbf{\tilde{M}}_{6}$, since there seems to be an allowed complex structure definable on $\tilde{M}_{6}$. Looking at it this way, the background M5-brane wraps an associative 3-cycle in $\tilde{M}_{6}$, in particular $Re(\Psi_{(3)})$ in our conventions, where $\Psi_{(3)}$ is the (3,0)-form $\Psi_{(3)} =  \xi^1 \wedge \xi^2 \wedge \xi^3$. This is consistent with the known fact that an M5-brane wrapping an associative 3-cycle in a Calabi-Yau 3-fold spanning $12389(10)$ has a calibrating form $\Upsilon = i H^{-1/6} \tilde{J}$~\cite{Martelli:2003ki,Fayyauddin:2005as}. 

Lastly, we would like to note that, in the first instance, one can probe the background with an M5-brane embedded holomorphically with respect to $J^{\prime}$ with the projection condition

\begin{equation}
\left(e^{i\phi}\hat{\Gamma}_{0ab} + e^{-i\phi}\hat{\Gamma}_{0\bar{a}\bar{b}}\right)\hat{\Gamma}_{89(10)}\epsilon = \epsilon .
\label{ProjpM5}
\end{equation}
Then we would find that, given the identity $\hat{\Gamma}_{0123456789(10)}\equiv 1$, there was a ``hidden" M2-brane embedded holomorphically with respect to $-J^{\prime\prime}$ with the projection condition

\begin{equation}
\left(e^{i(\phi-\pi /2)}\hat{\Gamma}_{0ab} + e^{-i(\phi - \pi /2)}\hat{\Gamma}_{0\bar{a}\bar{b}}\right)\epsilon = \epsilon .
\label{ProjhM2}
\end{equation}
The corresponding central charge for this brane would be 

\begin{equation}
\mp \int \left(   e^{-i(\phi - \pi /2)} dF^2 \wedge dG + e^{i(\phi - \pi /2)} d\bar{F}^2 \wedge d\bar{G} \right) .
\end{equation}
and the relevant term in the M5-brane central charge would become

\begin{equation}
\mp \int \left( \ldots + (e^{-i\phi} dF^2 \wedge dG + e^{i\phi} d\bar{F}^2 \wedge d\bar{G} ) \wedge dx^8 \wedge dx^9 \wedge dx^{(10)} \ldots \right) 
\end{equation}
which are the same results as before except that the rotation of complex structures is in the opposite direction to what we had previously. This would be interpreted as an anti-M2-brane (when compared to our original M2-brane probe example), for instance. It makes no qualitative difference to the answer though.

To help visualise the branes at some limit we can make a table. The singular limit of these wrapped M5-branes on holomorphic 2-cycles is given by orthogonally intersecting five-branes, which we lay out here to make the setup more transparent, see Table~\ref{C2table}. The worldvolume directions spanned by the M5-branes that source the background are indicated by $\otimes$, with the allowed probe branes having worldvolume directions denoted by $\odot$. We have drawn double vertical lines to point out the $\tilde{M}_{6}$ manifold spanned by $12389(10)$. We have also drawn single vertical lines to denote the $\mathbf{C}^{2}$ subspace, which contains probes wrapped on all three complex structures, as can be seen from the middle entries. This singular limit shows the probe M2-brane wrapped on $J^{\prime}$, and the ``hidden" M5-brane wrapped on $J^{\prime\prime}$. Also shown are the M2-branes corresponding to BPS vortices (which wrap a holomorphic 2-cycle in $\tilde{M}_{6}$) and $\ast$ D6 denotes the object that would correspond to a D6-brane in Type IIA string theory. 

\begin{table*}[htbp]
	\centering
		\begin{tabular}{|cc||ccc|cccc|ccc||} \hline
			 & $0$ & $1$ & $2$ & $3$ & $\Re (G)$ & $\Im (G)$ & $\Re (F^{2})$ & $\Im (F^{2})$ & 8 & 9  & $10$ \\ \hline
			 M5 & $\otimes$ & $\otimes$ & $\otimes$ & $\otimes$  & $\otimes$ & $\otimes$ & & & & & \\ 
			 M5 & $\otimes$ & $\otimes$ & $\otimes$ & $\otimes$  & & & $\otimes$ & $\otimes$ & & & \\ 
			 M2 & $\odot$ &  & & & $\odot$ & & $\odot$ & & & & \\
			 M2 & $\odot$ & $\odot$ & & &  & &  & & & & $\odot$ \\
			 M5 & $\odot$ & & $\odot$ & & $\odot$ & $\odot$ & &  & $\odot$ &  & $\odot$  \\
			 M5 & $\odot$ & & & & $\odot$ &  & & $\odot$ & $\odot$ & $\odot$ & $\odot$  \\ 
			 $\ast$ D6 & $\odot$ & $\odot$  & $\odot$ & $\odot$ & & & & & $\odot$ & $\odot$ & $\odot$  \\ \hline
		\end{tabular}
\caption{Some possible supersymmetric probe embeddings. The real and imaginary parts of our coordinates are denoted by $G= \Re (G) + i \Im (G)$.}
\label{C2table}
\end{table*}

\section{Central charges of an M5-brane wrapping a holomorphic 2-cycle in $\mathbf{C}^3$}
\label{N=1CentralCharge}

In this section we calculate the central charge for the last background we are examining, M5-branes wrapped on a holomorphic 2-cycle in $\mathbf{C}^3$, which is the supergravity dual of $\mathcal{N}=1$ MQCD supersymmetric gauge theories in four dimensions.

As opposed to the previous example, in which we considered a time-like vector $K$, this case corresponds to a background geometry with a null vector $K$. Since our background preserves four supercharges, the original $(Spin(7) \ltimes {\mathbb{R}}^{8}) \times \mathbb{R}$ isotropy group will be broken down, reducing the isometries of our background. Thus, our projection conditions will differ in structure from those in the last section, reflecting the different isometries of this geometry.

To start with, the projection condition satisfied by the background M5-brane wrapped on a holomorphic 2-cycle is given by

\begin{equation}
\hat{\Gamma}_{0123a\bar{b}}\epsilon = i \delta_{a\bar{b}}\epsilon,
\label{ProjBM5}
\end{equation}
where now we recall that $a,b=1,2,3$ (since $\mathbf{C}^3$ is defined in the $456789$ space). We also find that we can choose a compatible projection involving the $01$ directions of the form

\begin{equation}
\hat{\Gamma}_{01}\epsilon  =  \pm \epsilon
\label{Gamma01Proj}
\end{equation}
where either sign can be chosen. This is also equivalent to adding momentum along the $1$ direction, and is well known to break a further $\frac{1}{2}$ supersymmetry. Similarly, for the $23$ and $z^{1},z^{2},z^{3}$ spaces, we have a certain freedom to impose compatible projections. If we permit an arbitrary angle in the $23$ plane and an arbitrary phase for the $z^{1},z^{2},z^{3}$ space, we have

\begin{equation}
\left( e^{i\phi} \left(\alpha \hat{\Gamma}_{2} - \beta \hat{\Gamma}_{3} \right) \hat{\Gamma}_{z_{1}z_{2}z_{3}} + e^{-i\phi} \left(\alpha \hat{\Gamma}_{2} - \beta \hat{\Gamma}_{3} \right) \hat{\Gamma}_{\overline{z_{1}z_{2}z_{3}}} \right)\epsilon  =  \epsilon
\end{equation}
with the condition that $\alpha^2 + \beta^2 = 1$ and we can check this projector is also Hermitian, as is required. We may set $\alpha = \cos \theta$ and $\beta = \sin\theta$ at this point. Again, the equations for $1/8$-SUSY hold for arbitrary phase $\phi$. 

Finally, one should note that using the identity $\hat{\Gamma}_{0123456789y}\equiv 1$ we can show that our projections imply:

\begin{equation}
\hat{\Gamma}_{y}\epsilon = -\epsilon .
\label{GammayProj}
\end{equation}
These provide a set of independent, commuting projections which determine a unique spinor up to scale. The scale of the spinor in this case is fixed, again using the fact that $K$ is a Killing vector, to be $\epsilon^{\dagger}\epsilon  = \Delta = H^{-1/6}$.

As in the previous section, we can now proceed to calculate the non-trivial components of each form. For example, in this case we can easily show that the $K_{i} (i=2,3), K_{a}, K_{\overline{b}} (a,b=1,2,3)$ and $K_{y}$ components vanish since, for example,
$$
\begin{array}{ccccccccc}
\hat{K}_{2} & = & \overline{\epsilon} \hat{\Gamma}_{2} \epsilon & = & \pm \overline{\epsilon} \hat{\Gamma}_{2} \hat{\Gamma}_{01}\epsilon & = & \pm \overline{\epsilon} \hat{\Gamma}_{012} \epsilon & = & 0 
\end{array}
$$
where in the second step we have used the $\hat{\Gamma}_{01}$ projection condition~(\ref{Gamma01Proj}), and in the last step the fact that the three-form vanishes identically.

In analogy with the previous $\mathcal{N}=2$ case, and for the purposes of calculation, we can define the coordinates $F^2,E^2$ which are locally perpendicular to the brane, and $G$ which is locally parallel to the brane. The function $H$ should thus be harmonic in $F^2,E^2$ and $y$. The metric in these coordinates takes the form

\begin{equation}
ds^2 = H^{-1/3} dx_{(1,3)}^2 + 2 H^{1/6} \left( H^{-1/2}\left\vert dG \right\vert^2 + H^{1/2}\left\vert dF^2 \right\vert^2 + H^{1/2}\left\vert dE^2 \right\vert^2 \right) + H^{2/3} dy^2 
\label{N1FGKmetric}
\end{equation}
One can check that the equations of motion are satisfied with this metric. 
Computing the rest of the forms we find the following:

\begin{eqnarray}
K & = & -H^{-1/3} \left( dt \mp dx^1 \right) \\
\Omega &=& \left( dt \mp dx^1 \right) \wedge dy \\
\Sigma &=& \mp i H^{-1/2} g_{M\bar{N}}  \left(dt \mp dx^1 \right) \wedge dx^2 \wedge dx^3 \wedge dz^{M} \wedge dz^{\bar{N}} \nonumber \\
& & +  \left(dt \mp dx^1 \right) \wedge (\omega \wedge \omega) \nonumber \\
& & - \frac{H^{-1/2}}{2} \left(dt \mp dx^1 \right) \wedge e^{\mp i \theta} \left( dx^2 \mp i dx^3 \right) \wedge e^{-i\phi} \Psi_{(3)} \nonumber \\
& &  - \frac{H^{-1/2}}{2}  \left(dt \mp dx^1 \right) \wedge e^{\pm i \theta} \left( dx^2 \pm i dx^3 \right) \wedge e^{i\phi} \bar{\Psi}_{(3)} .
\end{eqnarray}
We have denoted the holomorphic three-form as $\Psi_{(3)} = e^{z^{1}} \wedge e^{z^{2}} \wedge e^{z^{3}}$. We shall write the vielbein $e^{z^{1}}$ to avoid confusion with the $x^1 $ part of the metric. 

Furthermore, if we choose for convenience the bottom signs in the lines above and define the complex co-ordinate $\lambda = x^2 + i x^3$, then we can re-write it in the following way:

\begin{eqnarray}
\Sigma_{K(\lambda\Psi + \bar{\lambda}\bar{\Psi})} &=& - \frac{H^{-1/2}}{2} \left(dt \mp dx^1 \right) \wedge e^{i (\theta - \phi)} \left[ d\lambda \wedge \Psi_{(3)} \right] \nonumber \\
& &  - \frac{H^{-1/2}}{2}  \left(dt \mp dx^1 \right) \wedge e^{- i (\theta - \phi)} \left[ d\bar{\lambda} \wedge \bar{\Psi}_{(3)} \right] .
\end{eqnarray}
In order to find the constraint on $\theta$ and $\phi$, it is useful to express this fully in terms of vielbeins and substitute in for $K$. We find 

\begin{equation}
\Sigma_{K(\Psi_{(4)} + \bar{\Psi}_{(4)})} = 1/2 \left[ K \wedge e^{i (\theta - \phi)} \Psi_{(4)}  + K \wedge e^{- i (\theta - \phi)} \bar{\Psi}_{(4)} \right] .
\end{equation}
We have defined the holomorphic four-form $\Psi_{(4)} =  e^\lambda \wedge \Psi_{(3)}$ in the enlarged $\mathbf{R}^{7} \times S^1$ subspace of the M-theory vacuum. Since this five-form $\Sigma$ should be real, and that it is determined uniquely~\cite{Hackett-Jones:2003vz} given our projections conditions, we find that 

\begin{equation}
\Sigma_{K\wedge Re(\Psi_{(4)})} = + K \wedge Re \left( e^{i(\theta - \phi)} \Psi_{(4)}  \right)  ,
\end{equation}
which is of the same form as in~\cite{Hackett-Jones:2003vz} and $Re \left( \Psi_{(4)}  \right) \subset \Phi_{(4)}$, with $\Phi_{(4)}$ the Cayley four-form. This implies that for any two supersymmetric brane probes of this type, the relative angle $\theta$ between them in the $\lambda\bar{\lambda}$-plane should be equal to the angle $\phi$ in the $89$-plane.

Proceeding in the same manner as before, we now need to calculate the six- and three-form potentials which are necessary to find the central charge. Once more the six-form is easy to recover since it is gauge equivalent to the calibration bound satisfied by the background M5-brane~(\ref{CalboundC3}). If we make sure to include the right asymptotic conditions and contract with $K$ we find that

\begin{equation}
\imath_{K}C = \pm i (H^{-1/2} g_{M\bar{N}} - \delta_{M\bar{N}})  \left(dt \mp dx^1 \right) \wedge dx^2 \wedge dx^3 \wedge dz^{M} \wedge dz^{\bar{N}} .
\end{equation} 
We can obtain the three-form magnetic potential in the same manner as before, defining the appropriate integral. We find that 

\begin{eqnarray}
\Omega \wedge F &=& \left( dt \mp dx^1 \right) \wedge dy \wedge \partial_{y}(\omega\wedge\omega) \nonumber \\
& = & \left( dt \mp dx^1 \right) \wedge d(\omega\wedge\omega) 
\end{eqnarray}
which gives us an expression for the three-form potential of the form

\begin{eqnarray}
\Omega \wedge A &=& -  \left(dt \mp dx^1 \right) \wedge (\omega \wedge \omega) \nonumber \\
& = & \left(dt \mp dx^1 \right) \wedge (g_{M[\bar{N}}g_{P\bar{Q}]} - \delta_{M\bar{N}}\delta_{P\bar{Q}}) dz^M \wedge dz^{\bar{N}} \wedge dz^P \wedge dz^{\bar{Q}}  . 
\end{eqnarray}
When combining this term with the second term from the expression for $\Sigma$, we find, after some cancellations, that we are left with something of the form $-\delta_{M\bar{N}}\delta_{P\bar{Q}}(dt \mp dx^1) \wedge dz^M \wedge dz^{\bar{N}} \wedge dz^P \wedge dz^{\bar{Q}}$. Defining the (1,1)-form on the Hermitian manifold $M_{6} \subset \mathbf{C}^{3}$ by $J_{f} = i\delta_{i\bar{j}} dz^i \wedge dz^{\bar{j}}$ and noting that the flat holomorphic three-form is given by $\Psi_{(3)f} = H^{-1/2}\Psi_{(3)}$, as before, we can simplify this expression somewhat. We also note that, as we saw in the previous example, the contraction $\imath_{K} A$ also vanishes for this background. We can now compile all the terms that make up the central charge, which yields

\begin{eqnarray}
\int K^{M}P_{M} & \geq & \mp \int \left(  \imath_{K}C + \Sigma + (A + dB) \wedge \Omega \right) \nonumber \\
& \geq & \mp \int \Bigl( \mp i \left(dt \mp dx^1 \right) \wedge dx^2 \wedge dx^3 \wedge J_{f} \Bigr. \nonumber \\
& & \left. +(dt \mp dx^1) \wedge J_{f} \wedge J_{f} \right. \nonumber \\
& & - \Bigl. \frac{1}{2} \left(dt \mp dx^1 \right) \wedge e^{i\theta} d\lambda \wedge e^{-i\phi} H^{-1/2} \Psi_{(3)}  \Bigr. \nonumber \\
& &  - \Bigl. \frac{1}{2}  \left(dt \mp dx^1 \right) \wedge e^{- i \theta} d\bar{\lambda} \wedge e^{i \phi} H^{-1/2} \bar{\Psi}_{(3)} \Bigr. \nonumber \\ 
& & \Bigl. + dB \wedge \Omega \Bigr) .  
\end{eqnarray}
So again we see that all the terms actually combine to give us the flat-space supersymmetry algebra, in the form of Eq. $(12)$ in~\cite{Hackett-Jones:2003vz}. Namely, the expression for the central charges takes the form

\begin{equation}
\left(dt \mp dx^1 \right) \wedge \Phi_{(4f)} + dB \wedge \Omega
\end{equation}
where $\Phi_{(4f)}$ is the flat-space Cayley four-form. 

The first possibility allowed by the central charge, for a suitable embedding, is the obvious case of a parallel probe brane. The next term allows for a probe wrapped on a holomorphic 4-cycle in $\mathbf{C}^3$. Of more interest are the next two terms. We have written them in a suggestive manner which we will explain shortly. These central charges are equivalent to a probe M5-brane wrapping a Cayley calibrated 4-cycle in some manifold $M_{8}$. This has a natural interpretation as intersecting MQCD domain walls preserving $1/16$ of the overall supersymmetry. They are thus $1/2$ BPS states of the worldvolume theory. We can further add some momentum along the $01$ directions (or along the null Killing vector) which again breaks half the supersymmetries leaving us with $1/32$. The domain wall interpretation can be illustrated by the sequence

\begin{equation}
\mathbf{R}^{(1,1)} \times Re \left( \Psi_{(4)}  \right) \rightarrow \mathbf{R}^{(1,2)} \times \Psi_{(3)} \rightarrow \mathbf{R}^{(1,3)} \times \Sigma_{2}
\end{equation}
which was made explicit in our construction. The 4-cycle calibrated by $Re \left( \Psi_{(4)} \right)$ contains a line in the $23$ ($\lambda \bar{\lambda}$) plane and also a 3-cycle calibrated by $\Psi_{(3)}$, where $\Psi_{(3)}$ is an associative 3-cycle. These are the individual domain walls. As one moves away from them, the space should change to $\mathbf{R}^{(1,3)} \times \Sigma_{2}$, representing the different vacua of the theory.

We can now justify this argument as follows. From the classification of eleven dimensional supergravity with a background null Killing spinor and flux~\cite{Cariglia:2004ym}, we can see that this is indeed the case. We started out with a background of an M5-brane wrapped on a 2-cycle in $\mathbf{C}^{3}$. We can see that since this preserves $1/8$ supersymmetry, this background has a $(SU(3) \ltimes {\mathbb{R}}^{6}) \times {\mathbb{R}}^3$ structure, as we would expect. This confirms, as before, that we are dealing with a complex manifold on which we can define a holomorphic three-form. In the singular limit, these three M5-branes will represent the vacua of our domain wall configuration.

We then saw, from the central charge result, that this background allows for a $1/2$ BPS M5-brane probe with worldvolume $\mathbf{R}^{(1,1)} \times Re \left( \Psi_{(4)}  \right)$, which was interpreted as a BPS intersecting domain wall on the worldvolume theory. This configuration now preserves $1/16$ supersymmetry. We can see that we were coherent in claiming that this M5-brane wraps a Cayley calibrated 4-cycle since we now have a $(SU(4) \ltimes {\mathbb{R}}^{8}) \times \mathbb{R}$ structure according to the classification. Finally, adding momentum along the Killing direction $01$ breaks a further $1/2$ supersymmetry, and accordingly, since we only have one supersymmetry generator left in our background, we have recovered the expected $(Spin(7) \ltimes {\mathbb{R}}^{8}) \times \mathbb{R}$ structure. So the geometrical description fits in rather nicely with the field theory interpretation.

Furthermore, our constraint on the $\theta$ and $\phi$ phase spaces has a simple interpretation. The vector representing the intersection angle of the domain walls in the $23$ space should be the same magnitude as the vector representing the angle between them in the `electric/magnetic' charge space. In our construction this was the $89$ space. This implies that, for example, we would have $2468 \rightarrow -3469$, which agrees with the standard form of the Cayley four-form. This agrees, up to conventions, with the results of~\cite{Gauntlett:2000yv}.

Lastly, we can calculate the tension of these domain walls. From the way we wrote the term, it is natural to conclude that it is given by

\begin{equation}
T_{DM} = \left\vert e^{-i\phi} \int H^{-1/2} \Psi_{(3)} \right\vert .
\end{equation}
This is consistent with the recent result~\cite{Fayyazuddin:2005ds} where it was found that $H^{-1/2} \Psi_{(3)}$ corresponds to a calibrating form in the geometry, and is thus closed. Therefore the integral represents a topological charge and we can conclude that these domain walls we have constructed are stable and have finite tension. 

The singular limit of these wrapped M5-branes on holomorphic 2-cycles is given by orthogonally intersecting five-branes, which we lay out here to make the setup more transparent, see Table~\ref{C3table}. The worldvolume directions spanned by the M5-branes that source the background are indicated by $\otimes$, with the allowed probe branes having worldvolume directions denoted by $\odot$. We have drawn double lines to separate the $\mathbf{R}^{7} \times S^1$ subspace from the rest. The fourth entry denotes a probe wrapped on a holomorphic 4-cycle in $\mathbf{C}^3$. The bottom two entries clearly show an example of a Cayley calibrated 4-cycle. Within this space, delimited by a single vertical line, is the $\mathbf{C}^{3}$ subspace, which contains the associative 3-cycle as can also be seen from the bottom two entries. This singular limit shows two domain walls intersecting along the $01$ direction (with momentum running along this direction) and making a $\pi /2$ angle to each other in the $23$ plane and $- \pi /2$ angle in the $89$ plane. 

\begin{table*}[htbp]
	\centering
		\begin{tabular}{|ccc||cc|cccccc||c|} \hline
			 & $0$ & $1$ & $2$ & $3$ & $\Re (G)$ & $\Im (G)$ & $\Re (F^{2})$ & $\Im (F^{2})$ & $\Re (E^{2})$ & $\Im (E^{2})$ & $10$ \\ \hline
			 M5 & $\otimes$ & $\otimes$ & $\otimes$ & $\otimes$  & $\otimes$ & $\otimes$ & & & & & \\ 
			 M5 & $\otimes$ & $\otimes$ & $\otimes$ & $\otimes$  & & & $\otimes$ & $\otimes$ & & & \\ 
			 M5 & $\otimes$ & $\otimes$ & $\otimes$ & $\otimes$  & & & & & $\otimes$ & $\otimes$ & \\
			 M5 & $\odot$ & $\odot$ &  &  & $\odot$ & $\odot$ & & & $\odot$ & $\odot$ & \\
			 M5 & $\odot$ & $\odot$ & $\odot$ & & $\odot$ & & $\odot$ & & $\odot$ & & \\
			 M5 & $\odot$ & $\odot$ &  & $\odot$ & $\odot$ & & $\odot$ & &  & $\odot$ & \\ \hline
		\end{tabular}
\caption{Some possible supersymmetric probe embeddings. The real and imaginary parts of our coordinates are denoted by $G= \Re (G) + i \Im (G)$.}
\label{C3table}
\end{table*}

\section{Discussion}

In this paper we have calculated the central charges on the worldvolume of M2- and M5-brane probes in a background of M5-branes which are wrapped on 2-cycles in $\mathbf{C}^{2}$ and $\mathbf{C}^{3}$. This has revealed what supersymmetric M-brane probes of the eleven-dimensional supergravity solutions~\cite{Fayyazuddin:1999zu,Brinne:2000fh,Brinne:2000nf} are allowed. These probes have revealed interesting features about the corresponding $\mathcal{N}=2$ and $\mathcal{N}=1$ field theories.

In the case of the background sourced by M5-branes wrapping a holomorphic 2-cycle in $\mathbf{C}^{2}$, we found the known cases of a parallel M5-brane and the M2-brane which gives the mass of the BPS states of the four-dimensional $\mathcal{N}=2$ field theory. In particular, BPS monopoles and vortices where found. In addition, there was the case of the ``hidden" M5-brane, wrapped on the remaining complex structure of the hyper-K\"{a}hler manifold and extended along $89(10)$. This was related to the structure groups of reduced supersymmetry of M-theory vacua. Since this M5-brane wraps a calibrated cycle in the manifold $\tilde{M}_{6}$, it has a similar worldvolume interpretation to the M2-brane realising the BPS states of the theory, except with an extra volume modulus. Due to the possible boundary conditions when reduced to Type IIA Hanany-Witten configurations, these BPS vortices and monopoles may have finite tension.

In the case of the background sourced by M5-branes wrapping a holomorphic 2-cycle in $\mathbf{C}^{3}$, we found the interesting possibility of M5-branes wrapping Cayley calibrated 4-cycles, which changed into M5-branes wrapping associative 3-cycles. This was interpreted as a system of intersecting BPS domain walls. A constraint on the angle of intersection and the angle in charge space was derived. Also, the tension of the domain walls was found to be the integral of a calibrating form in the geometry. A discussion on null structure groups of M-theory vacua with flux showed these arguments to be consistent, providing a physical realisation in terms of M5-branes.

These examples provide a physical realisation of structure groups of M-theory vacua with flux, providing a more intuitive picture in terms of the geometry. It would be interesting to construct more examples of this kind.

\section{Acknowledgements} 

JSL would like to thank Maria Dolores Loureda for hospitality and the Sir Richard Stapley Educational Trust Fund for partial support.

\bibliographystyle{utphys}

\bibliography{references}

\end{document}